\pdfoutput=1
\documentclass[11pt]{article}

\usepackage{EMNLP2023}

\usepackage{times}
\usepackage{latexsym}

\usepackage[T1]{fontenc}
\usepackage[utf8]{inputenc}

\usepackage{microtype}

\usepackage{inconsolata}

\usepackage{booktabs} % For formal tables
\usepackage{adjustbox}
\usepackage{algorithm, algpseudocode}
\usepackage{amsmath}
\usepackage{graphics}
\usepackage{epsfig}
\usepackage{graphicx}
\usepackage{xcolor}
\usepackage[skip=5pt]{caption}
\usepackage{subfigure}
\usepackage{balance}
\usepackage{multirow}
\usepackage{mathrsfs}
\usepackage{acronym}
\usepackage{placeins}
\usepackage{xcolor}
\usepackage[inline]{enumitem}

\usepackage{url}
\usepackage{xspace}
\usepackage{bbm}

\acrodef{NER}{named entity recognition}
\acrodef{PLTR}{prompt learning with type-related features}
\acrodef{OOD}{out-of-domain}
\acrodef{TRF}{type-related feature}

\newcommand{\header}[1]{\vspace*{1mm}\noindent\textbf{#1.}}

\title{Generalizing Few-Shot Named Entity Recognizers to Unseen Domains \\ with Type-Related Features}

\author{
Zihan Wang$^{1,2}$, Ziqi Zhao$^1$, Zhumin Chen$^1$, Pengjie Ren$^1$, \\
\textbf{Maarten de Rijke}$^2$ and \textbf{Zhaochun Ren}$^3$\thanks{\hspace{1mm} Corresponding author.}\\
$^1$Shandong University, Qingdao, China \\ 
$^2$University of Amsterdam, Amsterdam, The Netherlands\\ 
$^3$Leiden University, Leiden, The Netherlands \\ 
\texttt{\{zihanwang.sdu,ziqizhao.work\}@gmail.com}, \texttt{chenzhumin@sdu.edu.cn}\\
\texttt{jay.ren@outlook.com}, \texttt{m.derijke@uva.nl}, \texttt{z.ren@liacs.leidenuniv.nl}\\ 
}

\begin{document}
\maketitle
\begin{abstract}
Few-shot \acf{NER} has shown remarkable progress in identifying entities in low-resource domains.
However, few-shot \ac{NER} methods still struggle with \ac{OOD} examples due to their reliance on manual labeling for the target domain.
To address this limitation, recent studies enable generalization to an unseen target domain with only a few labeled examples using data augmentation techniques.
Two important challenges remain:
First, augmentation is limited to the training data, resulting in minimal overlap between the generated data and \ac{OOD} examples. 
Second, knowledge transfer is implicit and insufficient, severely hindering model generalizability and the integration of knowledge from the source domain.
In this paper, we propose a framework, \acfi{PLTR}, to address these challenges.
To identify useful knowledge in the source domain and enhance knowledge transfer, \ac{PLTR} automatically extracts entity \acp{TRF} based on mutual information criteria. 
To bridge the gap between training and \ac{OOD} data, \ac{PLTR} generates a unique prompt for each unseen example by selecting relevant \acp{TRF}. 
We show that \ac{PLTR} achieves significant performance improvements on in-domain and cross-domain datasets. 
The use of \ac{PLTR} facilitates model adaptation and increases representation similarities between the source and unseen domains.\footnote{Our code is available at \url{https://github.com/WZH-NLP/PLTR}.}
\end{abstract}
\section{Introduction}
\label{sec:intro}
Named entity recognition (NER) aims to detect named entities in natural languages, such as locations, organizations, and persons, in input text \cite{DBLP:conf/sigir/00010WLSX22,sang2003introduction,jointNER}. 
This task has gained significant attention from both academia and industry due to its wide range of uses, such as question answering and document parsing, serving as a crucial component in natural language understanding \cite{nadeau2007survey, lstm-cnn-crf,lan,luke}. 
The availability of labeled data for NER is limited to specific domains, leading to challenges for generalizing models to new domains \cite{DBLP:conf/acl/LeeKTAF0MSPR22,cui-etal-2021-template,ma2021template}.

\begin{figure*}
  \centering
  \subfigure[Average similarities between pairs of sentences.]{
    \label{fig:exp_similarities}
    \includegraphics[width=0.48\linewidth]{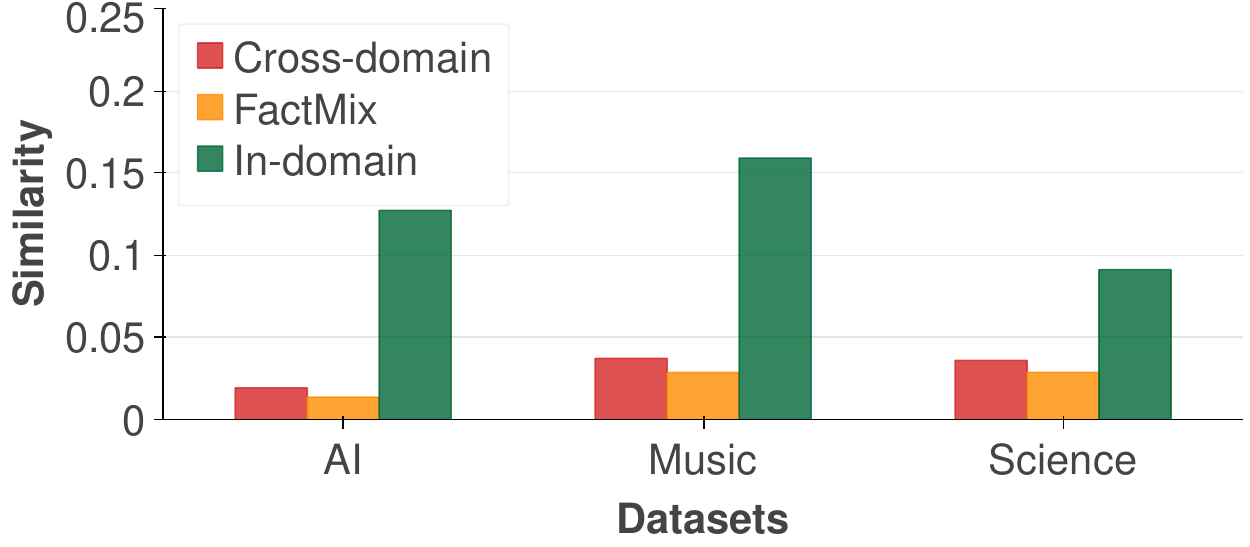}}
   \subfigure[Type-related features in the source domain.]{
    \label{fig:exp_knowledge}
    \includegraphics[width=0.48\linewidth]{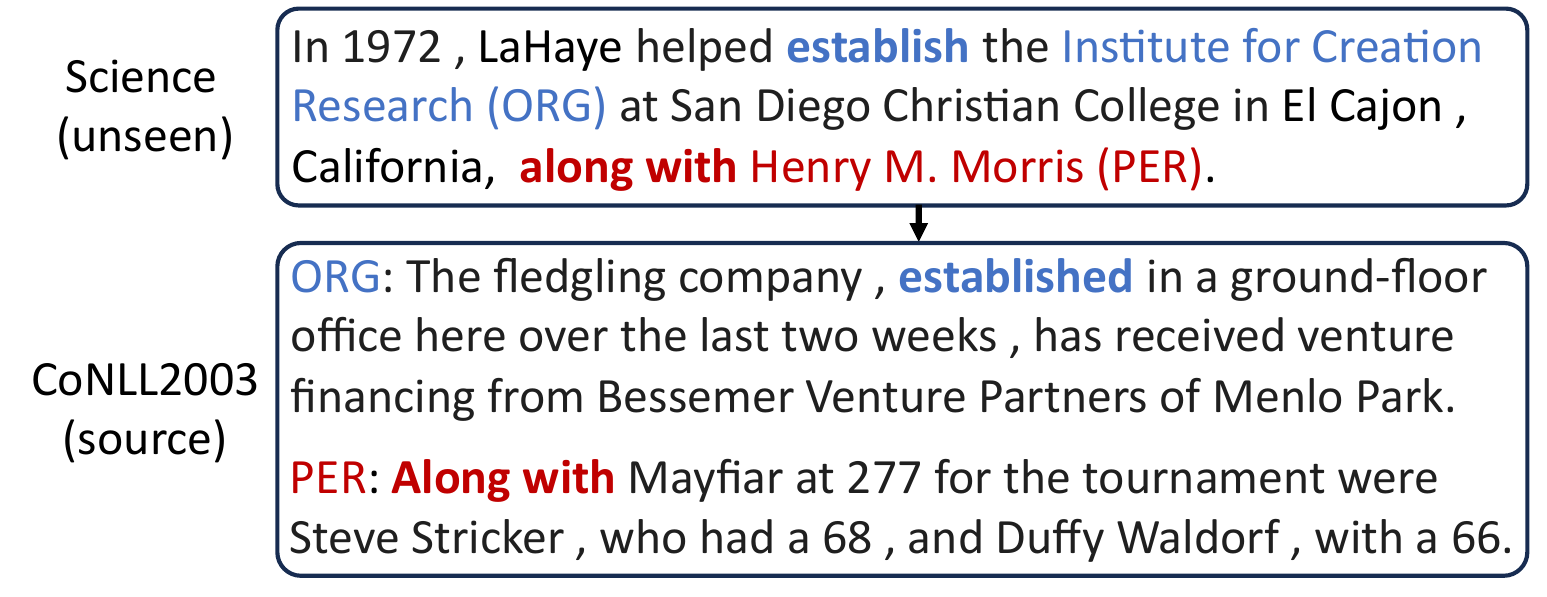}}
    \caption{(a) Average SBERT similarities~\citep{DBLP:conf/emnlp/ReimersG19} between pairs of sentences that contain the same type of entities. The source domain dataset is CoNLL2003~\citep{sang2003introduction}; the target domain datasets include AI, Music, and Science~\citep{liu2021crossner}. In the ``Cross-domain'' setting, one sentence is from the source domain and the other is from the target domain. In the ``FactMix'' setting, one sentence is from augmented data by FactMix~\citep{DBLP:conf/coling/YangYCGZ22}, and the other is from the target domain. In the ``In-domain'' setting, both sentences are from the target domain. (b) Examples of type-related features in the source domain.}
  \label{fig:examples}
\end{figure*}

To overcome this issue, recent research focuses on enabling models to effectively learn from a few labeled examples in new target domains \cite{DBLP:conf/acl/LeeKTAF0MSPR22,ma2021template,das2021container,chen2022few,wangusb,wang2022freematch} or on exploring data augmentation techniques, leveraging automatically generated labeled examples to enrich the training data \citep{zeng-etal-2020-counterfactual}.
However, these methods still require manual labeling for target domains, limiting their applicability in zero-shot scenarios with diverse domains.

Recently, \citet{DBLP:conf/coling/YangYCGZ22} have explored a new task, \emph{few-shot cross-domain} \ac{NER}, aiming to generalize an entity recognizer to unseen target domains using a small number of labeled in-domain examples.
To accomplish this task, a data augmentation technique, named FactMix, has been devised.
FactMix generates semi-fact examples by replacing the original entity or non-entity words in training instances, capturing the dependencies between entities and their surrounding context. 
Despite its success, FactMix faces two challenges: 

\header{Augmentation is limited to the training data}
Since the target domain is not accessible during training, FactMix exclusively augments the training data from the source domain. 
As a result, there is minimal overlap between the generated examples and the test instances at both the entity and context levels. 
For instance, only 11.11\% of the entity words appear simultaneously in both the generated data (by FactMix) and the AI dataset (target domain).
At the context level, as demonstrated in Fig.~\ref{fig:exp_similarities}, the average sentence similarity between the augmented instances and the test examples is remarkably low.
These gaps pose severe challenges in extrapolating the model to \ac{OOD} data.
To address this problem, we incorporate natural language prompts to guide the model during both training and inference processes, mitigating the gap between the source and unseen domains. 

\header{Knowledge transfer is implicit and insufficient}
Intuitively, better generalization to unseen domains can be accomplished by incorporating knowledge from the source domain~\citep{DBLP:journals/tacl/Ben-DavidOR22}.
However, in FactMix, the transfer of knowledge from the source domain occurs implicitly at the representation level of pre-trained language models.  
FactMix is unable to explicitly identify the \acfp{TRF}, i.e., tokens strongly associated with entity types, which play a crucial role in generalization.
E.g., as illustrated in Fig.~\ref{fig:exp_knowledge}, the words ``established'' and ``along with'' exhibit a close relationship with organization and person entities, respectively, in both domains. 
This knowledge can greatly assist in recognizing organizations and persons in the target domain.

To tackle this limitation, we introduce mutual information criteria to extract informative \acp{TRF} from the source domain.
Furthermore, we construct a unique prompt for each unseen instance by selecting relevant \acp{TRF}. Intuitively, these generated prompts serve as distinctive signatures, linking unfamiliar examples to the knowledge within the source domain.

\begin{figure*}[t]
    \centering
    \includegraphics[width=\linewidth]{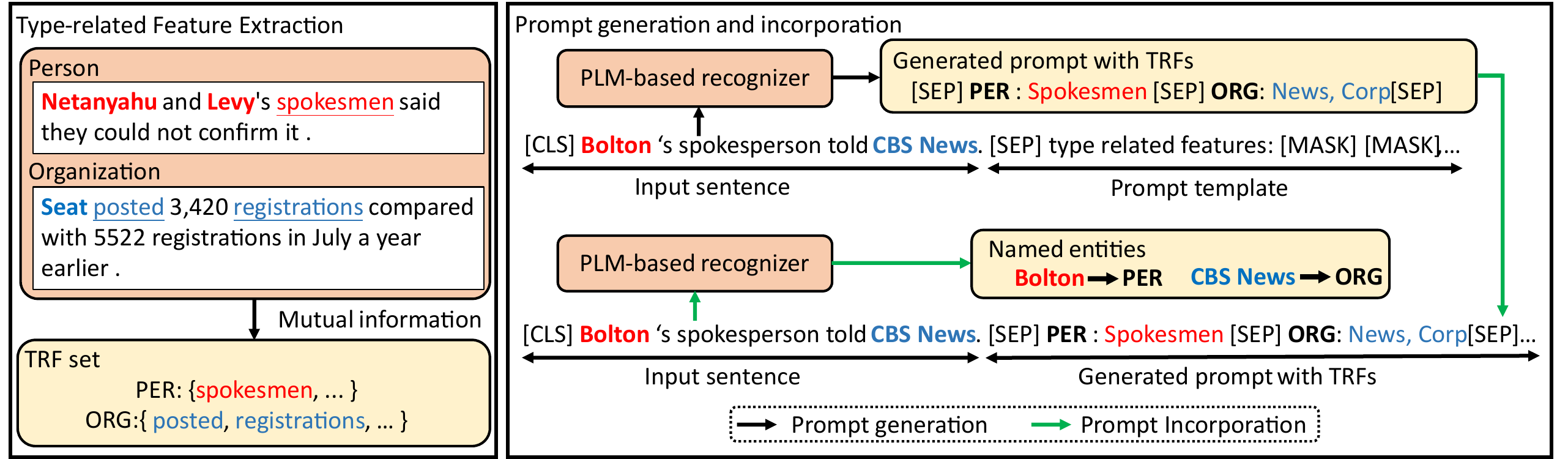}
    \caption{An overview of \ac{PLTR}. \ac{PLTR} has two main phases: \begin{enumerate*}[label=(\roman*), leftmargin=*,nosep]
    \item type-related feature extraction, and
    \item prompt generation and incorporation.
\end{enumerate*} Besides, we utilize a multi-task training strategy to enable parameter sharing between prompt generation and incorporation.}
    \label{fig:framework}
\end{figure*}

\header{Contributions}
In this paper, we present a framework, named \acfi{PLTR} for few-shot cross-domain \ac{NER}, to effectively leverage knowledge from the source domain and bridge the gap between training and unseen data.
As Fig.~\ref{fig:framework} shows, \ac{PLTR} is composed of two main phases:
\begin{enumerate*}[label=(\roman*), leftmargin=*,nosep]
    \item type-related feature extraction, and
    \item prompt generation and incorporation.
\end{enumerate*}
To identify valuable knowledge in the source domain, \ac{PLTR} uses mutual information criteria to extract entity \acfp{TRF}. 
\ac{PLTR} implements a two-stage framework to mitigate the gap between training and \ac{OOD} data.
Firstly, given a new example, \ac{PLTR} constructs a unique sequence by selecting relevant \acp{TRF} from the source domain. 
Then, the constructed sequences serve as prompts for performing entity recognition on the unseen data.
Finally, a multi-task training strategy is employed to enable parameter sharing between the prompt generation and entity recognition. 
Similar to FactMix, \ac{PLTR} is a fully automatic method that does not rely on external data or human interventions. 
\ac{PLTR} is able to seamlessly integrate with different few-shot \ac{NER} methods, including standard fine-tuning and prompt-tuning approaches.

In summary, our contributions are:
\begin{enumerate*}[label=(\roman*),nosep,leftmargin=*]
    \item to the best of our knowledge, ours is the first work to study prompt learning for few-shot cross-domain \ac{NER};
    \item we develop a mutual information-based approach to identify important entity type-related features from the source domain;
    \item we design a two-stage scheme that generates and incorporates a prompt that is highly relevant to the source domain for each new example, effectively mitigating the gap between source and unseen domains; and 
    \item experimental results show that our proposed \ac{PLTR} achieves state-of-the-art performance on both in-domain and cross-domain datasets.
\end{enumerate*}

\section{Related work}
\label{sec:related work}
\textbf{Cross-domain NER.} The task of cross-domain \ac{NER} aims to transfer NER models across diverse text styles \citep{pan2013transfer, liu2021crossner,chen2021data,lee-etal-2018-transfer,jointNER,jia2019cross,jia2020multi,DBLP:conf/acl/ZhengCM22,DBLP:conf/sigir/00010WLSX22,DBLP:conf/aaai/HuGLLCWC23,DBLP:conf/sigir/WangSRRCLLR21}. 
\citet{jointNER} train NER models jointly in the source and target domains, while \citet{jia2019cross} and \citet{jia2020multi} leverage language models for cross-domain knowledge transfer. 
\citet{DBLP:conf/sigir/00010WLSX22} introduce a modular learning approach that decomposes NER into entity span detection and type classification subtasks. However, these methods still rely on NER annotations or raw data in the target domain.

\header{Few-shot NER and prompt-based learning} 
Few-shot named entity recognition (NER) is the task of identifying predefined named entities using only a small number of labeled examples \citep{label-agnostic, nearest-neighbor-crf, das2021container, zeng-etal-2020-counterfactual,DBLP:conf/acl/MaLCZWGZGC23}. Various approaches have been proposed to address this task. For instance, \citet{huang-etal-2021-shot} investigate the effectiveness of self-training methods on external data using distance-based approaches, where the label of the nearest neighbors is copied. \citet{zeng-etal-2020-counterfactual} involves generating counterfactual examples through interventions to augment the original dataset.
Additionally, prompt-based learning, which has gained prominence in natural language processing, has also been applied to few-shot NER \citep{cui-etal-2021-template,ma2021template,DBLP:conf/acl/LeeKTAF0MSPR22,das2021container,DBLP:journals/corr/abs-2211-04337, DBLP:conf/coling/LiHC22,DBLP:journals/corr/abs-2308-14533,DBLP:conf/acl/FangWMXHJ23}. In particular,  \citet{das2021container} incorporate contrastive learning techniques with prompts to better capture label dependencies.
Furthermore, \citet{ma2021template} develop a template-free approach to prompt \ac{NER}, employing an entity-oriented objective.
Recently, several studies have conducted analyses of the performance of current large language models (LLMs), such as the GPT series~\citep{DBLP:conf/nips/BrownMRSKDNSSAA20,DBLP:journals/corr/abs-2303-08774}, in the context of the few-shot \ac{NER} task~\citep{DBLP:conf/emnlp/GutierrezMWCLS022,DBLP:journals/corr/abs-2305-14450,DBLP:journals/corr/abs-2306-09719}. 
Nevertheless, these investigations have revealed a substantial performance gap between recent LLMs and state-of-the-art methods.
Consequently, due to their high running costs and underwhelming performance, we do not consider recent LLMs as the basic model of our proposed framework (refer to Sec.~\ref{subsec:basic}).
As mentioned in Sec.~\ref{sec:intro}, previous few-shot \ac{NER} methods primarily focus on in-domain settings and require manual annotations for the target domain, which poses a challenge for generalizing to \ac{OOD} examples.

The field of \textbf{few-shot cross-domain learning} is inspired by the rapid learning capability of humans to recognize object categories with limited examples, known as rationale-based learning~\citep{DBLP:conf/nips/BrownMRSKDNSSAA20, Shen2021TowardsOG, chen2022few, baxter2000model, zhang2020every}. In the context of NER, \citet{DBLP:conf/coling/YangYCGZ22} introduce the few-shot cross-domain setting and propose a two-step rationale-centric data augmentation method, named FactMix, to enhance the model's generalization ability.

In this paper, we focus on few-shot cross-domain \ac{NER}. The most closely related work is FactMix~\citep{DBLP:conf/coling/YangYCGZ22}.
FactMix faces two challenging problems:
\begin{enumerate*}[label=(\roman*), leftmargin=*,nosep]
    \item augmentation is limited to the training data, and
    \item the transfer of knowledge from the source domain is implicit and insufficient.
\end{enumerate*}
In our proposed \ac{PLTR}, to identify useful knowledge in the source domain, mutual information criteria are designed for automatic \acf{TRF} extraction.
In addition,  \ac{PLTR} generates a unique prompt for each unseen example based on relevant TRFs, aiming to reduce the gap between the source and unseen domains.

\section{Preliminaries}

\subsection{Task settings}
\label{sec:settings}
A NER system takes a sentence \(\mathbf{x}=x_{1}, \ldots, x_{n}\) as input, where \(\mathbf{x}\) is a sequence of \(n\) words. It produces a sequence of NER labels \(\mathbf{y}=y_{1}, \ldots, y_{n}\), where each \(y_{i}\) belongs to the label set \(\mathcal{Y}\) selected from predefined tags \(\{B_t, I_t, S_t, E_t, O\}\). The labels \(B\), \(I\), \(E\), and \(S\) indicate the beginning, middle, ending, and single-word entities, respectively. The entity type is denoted by \(t\in\mathcal{T}=\{\text{PER}, \text{LOC}, \text{ORG}, \text{MISC},\ldots\}\), while \(O\) denotes non-entity tokens. The source dataset and out-of-domain dataset are represented by \(\mathcal{D}_\mathit{in}\) and \(\mathcal{D}_\mathit{ood}\), respectively.
Following~\citet{DBLP:conf/coling/YangYCGZ22}, we consider two settings in our task, the \textit{in-domain setting} and the \textit{\acf{OOD} setting}.
Specifically, we first train a model \(\mathcal{M}_{in}\) using a small set of labeled instances from \(\mathcal{D}_{in}\).
Then, for in-domain and \ac{OOD} settings, we evaluate the performance of \(\mathcal{M}_{in}\) on \(\mathcal{D}_{in}\) and \(\mathcal{D}_{ood}\), respectively.

\subsection{Basic models}
\label{subsec:basic}
Since our proposed \ac{PLTR} is designed to be model-agnostic, we choose two popular \ac{NER} methods, namely standard fine-tuning and prompt-tuning respectively, as our basic models.
As mentioned in Sec.~\ref{sec:related work}, due to their high costs and inferior performance on the NER task, we do not consider recent large language models (e.g., GPT series) as our basic models.

\header{Standard fine-tuning method}
We employ pre-trained language models (PLMs) such as BERT~\citep{devlin2018bert} and RoBERTa~\citep{liu2019roberta} to generate contextualized word embeddings. These embeddings are then input into a linear classifier with a softmax function to predict the probability distribution of entity types. The process involves feeding the input token \(x\) into the feature encoder \(\operatorname{PLM}\) to obtain the corresponding contextualized word embeddings $\mathbf{h}$:
\begin{equation}
\mathbf{h}=\operatorname{PLM}({x}),
\end{equation}
where \(\mathbf{h}\) represents the sequence of contextualized word embeddings derived from the pre-trained language models. To recognize entities, we optimize the cross-entropy loss \(\mathcal{L}_\mathit{NER}\) as:
\begin{equation}
    \mathcal{L}_\mathit{NER}= -\sum_{c=1}^{N} y_{o, c} \log \left(p_{o, c}\right),
\end{equation}
where \(N\) denotes the number of classes, \(y\) is a binary indicator (0 or 1) indicating whether the gold label \(c\) is the correct prediction for observation \(o\), and \(p\) is the predicted probability of \(c\) for \(o\).

\header{Prompt-tuning method}
The prompt-tuning method for \ac{NER} tasks involves the use of mask-and-infill techniques based on human-defined templates to generate label words. We adopt the recent EntLM model proposed by~\citet{ma2021template} as our benchmark for this method. First, a label word set \(\mathcal{V}_l\) is constructed through label word engineering, which is connected to the label set using a mapping function \(\mathcal{M}: \mathcal{Y} \rightarrow \mathcal{V}_l\). Next, entity tokens at entity positions are replaced with the corresponding label word \(\mathcal{M}(y_i)\). The resulting modified input is then denoted as \(\mathbf{x^{Ent}}=\left\{x_{1}, \ldots, \mathcal{M}\left(y_{i}\right), \ldots, x_{n}\right\}\). The language model is trained by maximizing the probability \(P\left(\mathbf{x^\mathit{Ent}} \mid \mathbf{x}\right)\). The loss function for generating the prompt and performing \ac{NER} is formulated as:
\begin{equation}
    \mathcal{L}_\mathit{NER}=-\sum_{i=1}^{N} \log P\left(x_{i}=x_{i}^\mathit{Ent} \mid \mathbf{x}\right),
\end{equation}
where \(N\) represents the number of classes. The initial parameters of the predictive model are obtained from PLMs.

\section{Method}
In this section, we present the two primary phases of the proposed \ac{PLTR} method, as depicted in Fig.~\ref{fig:framework}:
\begin{enumerate*}[label=(\roman*), leftmargin=*, nosep]
\item type-related feature extraction (see Sec.~\ref{subsec:extraction}), and
\item prompt generation and incorporation (see Sec.~\ref{subsec:prompt}).
\end{enumerate*}

\subsection{Type-related feature extraction}
\label{subsec:extraction}
As mentioned in Sec.~\ref{sec:intro}, \acfp{TRF}, which are tokens strongly associated with entity types, play a crucial role in the few-shot cross-domain \ac{NER} task. To extract these features, we propose a mutual information based method for identifying \acp{TRF} from the source domain. 
Here, we define $\mathcal{S}_{i}$ as a set that contains all sentences from the source domain where entities of the $i$-th type appear, and $\mathcal{S} \backslash \mathcal{S}_i$ as a set that contains sentences without entities of the $i$-th type.
In our method, we consider a binary variable that indicates examples (texts) from $\mathcal{S}_{i}$ as $1$, and examples from $\mathcal{S} \backslash \mathcal{S}_i$ as $0$. 
To find tokens closely related to $\mathcal{S}_{i}$, we first calculate the mutual information between all tokens and this binary variable, and then select the top $l$ tokens with the highest mutual information scores. However, the mutual information criteria may favor tokens that are highly associated with $\mathcal{S} \backslash \mathcal{S}_i$ rather than with $\mathcal{S}_{i}$.
Thus, we introduce a filtering condition as follows:
\begin{equation}
\label{eq:MI}
\frac{C_{\mathcal{S}\backslash \mathcal{S}_i}(\mathbf{w}_{m})}{ C_{\mathcal{S}_i}(\mathbf{w}_{m})} \leq \rho, \;\;  C_{\mathcal{S}_i}(\mathbf{w}_{m}) > 0, 
\end{equation}
where $C_{\mathcal{S}_i}(\mathbf{w}_{m})$ represents the count of the m-gram $\mathbf{w}_{m} = x_p,\ldots, x_{p+m-1}$ in $\mathcal{S}_i$. $C_{\mathcal{S}\backslash \mathcal{S}_i}(\mathbf{w}_m)$ represents the count of this m-gram $\mathbf{w}_{m}$ in all source domains except for $\mathcal{S}_i$, and $\rho$ is an m-gram frequency ratio hyperparameter.
By applying this criterion, we ensure that $\mathbf{w}_{m}$ is considered part of the \ac{TRF} set of $\mathcal{S}_{i}$ only if its frequency in $\mathcal{S}_{i}$ is significantly higher than its frequency in other entity types ($\mathcal{S}\backslash \mathcal{S}_i$). Since the number of examples in $\mathcal{S}_i$ is much smaller than the number of examples in $\mathcal{S} \backslash \mathcal{S}_i$, we choose $\rho \geq 1$ but avoid setting it to a large value. This allows for the inclusion of features that are associated with $\mathcal{S}_i$ while also being related to other entity types in the \ac{TRF} set of $\mathcal{S}_i$. 
In our experiments, we set $\rho=3$ and only consider 1-gram texts for simplicity.

Note that the type-related feature extraction module we designed is highly efficient with a computational complexity of $\rm{O} (|\mathcal{D}_\mathit{in}| \cdot l_\mathit{avg} \cdot |\mathcal{T}|)$, where $|\mathcal{D}_\mathit{in}|$, $l_\mathit{avg}$, and $\mathcal{T}$ represent the number of sentences in the training dataset, the average sentence length, and the entity type set, respectively. This module is able to compute the mutual information criteria in Eq.~\ref{eq:MI} for all entity types in $\mathcal{T}$ and each token by traversing the tokens in every training sentence just once.

\subsection{Prompt generation and incorporation}
\label{subsec:prompt}
To connect unseen examples with the knowledge within the source domain, we generate and incorporate a unique prompt for each input instance. This process involves a two-stage mechanism: first, relevant \acp{TRF} are selected to form prompts, and then these prompts are input into the PLM-based basic model for entity label inference.

\header{Automatic type-related feature selection}
Given an input sentence $\mathbf{x}$ and the extracted \ac{TRF} set $\mathcal{R}$, we formulate the selection of relevant \acp{TRF} as a cloze-style task for our PLM-based basic model $\mathcal{M}_{b}$ (refer to Sec.~\ref{subsec:basic}). Specifically, we define the following prompt template function $f(\cdot)$ with $K$ $\texttt{[MASK]}$ tokens:
\begin{equation}
\label{eq:prompt1}
\resizebox{6.8cm}{!}{%
$
\begin{aligned}
    &f(\mathbf{x}) = \\ &\text{``}\mathbf{x}\texttt{[SEP]}
    \text{type-related features:} \texttt{[MASK]...[MASK]}\text{''}.
\end{aligned}
$}
\end{equation}
By inputting $f(\mathbf{x})$ into $\mathcal{M}_{b}$, we compute the hidden vector $\mathbf{h}_{\texttt{[MASK]}}$ of $\texttt{[MASK]}$. 
Given a token $r \in \mathcal{R}$, we compute the probability that token $r$ can fill the masked position:
\begin{equation}
\label{eq:select}
\resizebox{6.8cm}{!}{%
$
\displaystyle
p(\texttt{[MASK]}=r|f(\mathbf{x}))) = \frac{\exp(\mathbf{r}\cdot \mathbf{h}_{\texttt{[MASK]}})}{\sum_{\tilde{r}\in\mathcal{R}}\exp(\mathbf{\tilde{r}}\cdot \mathbf{h}_{\texttt{[MASK]}})},
$}
\end{equation}
where $\mathbf{r}$ is the embedding of the token $r$ in the PLM $\mathcal{M}_{b}$.
For each $\texttt{[MASK]}$, we select the token with the highest probability as the relevant \ac{TRF} for $\mathbf{x}$, while discarding any repeating \acp{TRF}. 
For example, as illustrated in Fig. \ref{fig:framework}, for the sentence ``Bolton's spokesperson told CBS News.'', the most relevant \acp{TRF} include ``Spokesmen'', ``News'' and ``Corp''.

To train $\mathcal{M}_{b}$ for \ac{TRF} selection, we define the loss function $\mathcal{L}_{gen}$ as follows:
\begin{equation}
{
    \small
\begin{aligned}
  &\mathcal{L}_\mathit{gen} = \\
  &- \frac{1}{|\mathcal{D}_\mathit{in}|} \sum_{x\in \mathcal{D}_\mathit{in}}{\sum_{i=1}^{K}\log p(\texttt{[MASK]}_{i} = \phi(x, i)|f(\mathbf{x}))},
\end{aligned}}
\end{equation}
where $\phi(x,i)$ denotes the label for the $i$-th $\texttt{[MASK]}$ token in $\mathbf{x}$. To obtain $\phi(x)$, we compute the Euclidean distance between the PLM-based embeddings of each $r\in\mathcal{R}$ and each token in $\mathbf{x}$, selecting the top-$K$ features.
Note that our designed automatic selection process effectively filters out irrelevant \acp{TRF} for the given input sentence, substantially reducing human interventions in \ac{TRF} extraction (refer to Sec.~\ref{sec:analysis}). 

\header{Prompt incorporation}
To incorporate the entity type information into prompts, we generate a unique prompt given the selected relevant \acp{TRF} $\mathcal{R}'(\mathbf{x})\subseteq\mathcal{R}$ for input $\mathbf{x}$. 
This is achieved using the following prompt template function $f'(\mathbf{x})$:
\begin{equation}
\label{eq:prompt2}
\resizebox{6.8cm}{!}{%
$
\begin{aligned}
    & f'(\mathbf{x}) = \\
    & \text{``}\mathbf{x}\texttt{[SEP]} 
     t_{1}{:} \mathcal{R}'(\mathbf{x}, t_{1})
     \texttt{[SEP]...} 
     % t_{2}: \mathcal{R}'(\mathbf{x}, t_{2})\texttt{[SEP]}
   \texttt{[SEP]}t_{|\mathcal{T}|}{:} \mathcal{R}'(\mathbf{x}, t_{|\mathcal{T}|})\text{''},
\end{aligned}
$}%
\end{equation}
where $t_{i}\in \mathcal{T}$ is the entity type name (e.g., $\text{PER}$ or $\text{ORG}$). Given sentence $\mathbf{x}$, $\mathcal{R}'(\mathbf{x}, t_{i})\subseteq\mathcal{R}'(\mathbf{x})$ represents selected \acp{TRF} related to entity type $t_{i}$. 
Note that, if $\mathcal{R}'(\mathbf{x}, t_{i})=\emptyset$, the entity type name, and relevant \acp{TRF} $\mathcal{R}'(\mathbf{x}, t_{i})$ are excluded from $f'(\mathbf{x})$.
For example, as depicted in Fig. \ref{fig:framework}, the unique prompt $f'(\mathbf{x})$ corresponding to $\mathbf{x} = \text{``Bolton's spokesperson told CBS News.''}$ can be represented as follows:
\begin{equation}
\resizebox{6.8cm}{!}{%
$
\begin{aligned}
    &f'(\mathbf{x}) = \text{``Bolton's spokesperson told CBS News.}\\    &\texttt{[SEP]}\text{PER:Spokesmen}\texttt{[SEP]}\text{ORG:News, Corp}\text{''}.
\end{aligned}
$%
}
\end{equation}
Then, we input $f'(\mathbf{x})$ into $\mathcal{M}_{b}$ to recognize entities in the given sentence $\mathbf{x}$.

\subsection{Joint training}
\label{subsec:joint}
To enable parameter sharing between prompt generation and incorporation, we train our model using a multi-task framework. The overall loss function is defined as follows:
\begin{equation}
\label{eq:overall}
\mathcal{L} = \alpha \cdot \mathcal{L}'_\mathit{NER} + (1-\alpha) \cdot \mathcal{L}_\mathit{gen},
\end{equation}
where $\mathcal{L}'_\mathit{NER}$ denotes the normalized loss function for the NER task loss $\mathcal{L}_\mathit{NER}$ (refer to Sec.~\ref{subsec:basic}).  
$\alpha$ is the weight assigned to $\mathcal{L}'_\mathit{NER}$ with prompts as inputs. The weight $1-\alpha$ is assigned to the loss function $\mathcal{L}_\mathit{gen}$ for type-related feature selection.
In our experiments, we optimize the overall loss function using AdamW \citep{DBLP:conf/iclr/LoshchilovH19}.
Sec.~\ref{subsec:training} gives the detailed training algorithm of \ac{PLTR}.

\section{Experiments}
We aim to answer the following research questions:
\begin{enumerate*}[label=(RQ\arabic*),leftmargin=*,nosep]
\item Does \ac{PLTR} outperform state-of-the-art fine-tuning methods on the few-shot cross-domain  NER task? (Sec.~\ref{subs:fine-tuning})
\item Can \ac{PLTR} be applied to prompt-tuning NER methods? (Sec.~\ref{subs:prompt-tuning})
Micro F1 is adopted as the evaluation metric for all settings.
\end{enumerate*}

\begin{table}[t]
\centering
\footnotesize
\begin{tabular}{l l ccc c}
\toprule
 & \multicolumn{3}{c}{\textbf{\# Instances}} & \\ 
 \cmidrule{2-4}
\textbf{Dataset} & Train & Dev & Test &  \multicolumn{1}{l}{\textbf{Entity types}} \\ 
 \midrule
CoNLL2003                 & 14,987      & 3,466      & 3,684     & \phantom{1}4 \\
 OntoNotes & 59,924 & 8,528& 8,262 &18\\
\midrule
TechNews           & -          & -         & 2,000      & \phantom{1}4\\
AI                      & -          & -         & \phantom{0,}431      & 14\\
Literature              & -          & -         & \phantom{0,}416      & 12\\
Music                   & -          & -         & \phantom{0,}456      &13\\
Politics                & -          & -         & \phantom{0,}651      &\phantom{1}9\\
Science                 & -          & -         & \phantom{0,}543      &17\\ 
\bottomrule
\end{tabular}
\caption{Statistics of the datasets used.}
\label{tab:datasets}
\end{table}

\subsection{Datasets}
Detailed statistics of both in-domain and out-of-domain datasets are shown in Table~\ref{tab:datasets}. 

\header{In-domain dataset} We conduct in-domain experiments on the CoNLL2003 dataset~\citep{sang2003introduction}. 
It consists of text in a style similar to Reuters News and encompasses entity types such as person, location, and organization.
Additionally, to examine whether \ac{PLTR} is extensible to different source domains and entity types, we evaluate \ac{PLTR} using training data from OntoNotes~\citep{weischedel2013ontonotes} (refer to Sec.~\ref{sec:source}).
OntoNotes is an English dataset consisting of text from a wide range of domains and 18 types of named entities, such as Person, Event, and Date.

\header{Out-of-domain datasets} We utilize the \ac{OOD} dataset collected by~\citet{liu2021crossner}, which includes new domains such as AI, Literature, Music, Politics, and Science. 
The vocabulary overlaps between these domains are generally small, indicating the diversity of the out-of-domain datasets~\citep{liu2021crossner}. 
Since the model trained on the source domain dataset (CoNLL2003) can only predict person, location, organization, and miscellaneous entities, we assign the label \textit{O} to all unseen labels in the \ac{OOD} datasets.

\subsection{Experimental settings and baselines}
We compare \ac{PLTR} with recent baselines in the following two experimental settings: 

\header{Fine-tuning} Following~\citet{DBLP:conf/coling/YangYCGZ22}, we employ the standard fine-tuning method (Ori) based on two pre-trained models with different parameter sizes: BERT-base, BERT-large, RoBERT-base, and RoBERT-large. 
All backbone models are implemented using the transformer package provided by Huggingface.\footnote{\url{https://huggingface.co/models}} 
For fine-tuning the NER models in a few-shot setting, we randomly select 100 instances per label from the original dataset (CoNLL2003) to ensure model convergence. 
The reported performance of the models is an average across five training runs.

\header{Prompt-tuning} 
Similar to~\citet{DBLP:conf/coling/YangYCGZ22}, we adopt the EntLM model proposed by \citet{ma2021template} as the benchmark for prompt-tuning.
The EntLM model is built on the BERT-base or BERT-large architectures.
We conduct prompt-based experiments using a 5-shot training strategy~\citep{ma2021template}. 
Additionally, we select two representative datasets, TechNews and Science, for the \ac{OOD} test based on the highest and lowest word overlap with the original training domain, respectively.

\smallskip\noindent
Additionally, we include a recent data augmentation method CF~\citep{zeng-etal-2020-counterfactual} and the state-of-the-art cross-domain few-shot \ac{NER} framework FactMix~\citep{DBLP:conf/coling/YangYCGZ22} as baselines in both of the above settings.
Note that, we report the results of FactMix's highest-performing variant for all settings and datasets.

\subsection{Implementation details}
Following~\citet{DBLP:conf/coling/YangYCGZ22}, we train all models for 10 epochs and employ an early stopping criterion based on the performance on the development dataset. 
The AdamW optimizer~\citep{DBLP:conf/iclr/LoshchilovH19} is used to optimize the loss functions.
We use a batch size of 4, a warmup ratio of 0.1, and a learning rate of 2e-5. 
The maximum input and output lengths of all models are set to 256.
For \ac{PLTR}, we search for the optimal loss weight $\alpha$ from \{0.1, 0.25, 0.5, 0.75, 0.9\}. 
The frequency ratio hyperparameter $\rho$ is set to 3 for all domains.

\section{Experimental results}
\begin{table}[t]
\centering
\small
\setlength{\tabcolsep}{1.5mm}
%\resizebox{\linewidth}{!}{
\begin{tabular}{l cccc}
\toprule
  & \multicolumn{4}{c}{\textbf{In-domain Fine-tuning Results}}        \\ 
 \cmidrule{2-5} 
 \textbf{Backbone} & \textbf{Ori} & \textbf{CF} & \textbf{FactMix} & \textbf{PLTR} \\ 
\midrule
 BERT-base-cased & 54.03 & 77.71 & 80.10 &\textbf{82.05*} \\
 BERT-large-cased & 65.38 & 81.11 & 83.04 & \textbf{83.75*}\\
 RoBERTa-base & 48.53 & 82.74 & 85.33 &\textbf{86.40*} \\
 RoBERTa-large & 65.70 & 85.20 & 86.91 & \textbf{88.03*} \\ 
\bottomrule
\end{tabular}
%}
\caption{In-domain fine-tuning results (Micro F1) on CoNLL2003. \(*\) indicates a statistically significant difference (t-test, p\(<\)0.05) when compared to FactMix}
\label{tab:in-domain fine-tuning}
\end{table}

\begin{table*}[t]
\centering
\small
\begin{tabular}{@{~}l l cccc l cccc}
\toprule
  &  &  \multicolumn{4}{c}{\textbf{OOD Fine-tuning Results}} &  & \multicolumn{4}{c}{\textbf{OOD Fine-tuning Results}}\\ 
 \cmidrule{3-6} \cmidrule{8-11} 
 \textbf{Backbone} & \textbf{Dataset} & \textbf{Ori} & \textbf{CF} & \textbf{FactMix} & \textbf{PLTR} & \textbf{Dataset} & \textbf{Ori} &\textbf{CF} & \textbf{FactMix} & \textbf{PLTR} \\ 
 \hline
 BERT-base-cased  & \multirow{4}{*}{TechNews} &41.46  &61.20 &65.20 &\textbf{67.39*}  & \multirow{4}{*}{Music} & 10.46 & 19.33 & 19.49 &\textbf{23.86*}  \\
 BERT-large-cased & & 52.63 & 67.51 & 69.98 &\textbf{70.51*}  &  & 12.00 & 19.64 & 19.97 &\textbf{27.84*}  \\
 RoBERTa-base & & 44.88 & 71.83  &73.62 &\textbf{75.06*} &  &11.78 & 22.24 &23.75 &\textbf{30.52*}\\
 RoBERTa-large & & 51.76 &73.11 &74.89 & \textbf{75.14*}  &  &14.44 & 21.13 &22.93 &\textbf{30.26*} \\ 
\midrule             
 BERT-base-cased & \multirow{4}{*}{AI} & 15.88 &22.49 &24.67 &\textbf{28.41*} & \multirow{4}{*}{Politics}  & 21.38 &41.84 &43.60 &\textbf{44.97*}\\
 BERT-large-cased & & 18.62 & 26.00 &26.25 & \textbf{30.25*}  &  &29.77 &43.37 &43.84 & \textbf{45.85*} \\
 RoBERTa-base & & 18.63 &32.03 &32.09 &\textbf{33.87*} & &26.81 &44.12 &44.66 &\textbf{47.56*} \\
 RoBERTa-large & & 23.27 &28.76 &30.06 &\textbf{31.97*} & &28.56 &45.87 &45.05 &\textbf{48.35*}  \\ 
\midrule
 BERT-base-cased & \multirow{4}{*}{Literature} & 12.85  & 22.89 &25.70 &\textbf{27.39*} &\multirow{4}{*}{Science} & 12.41 &25.67 &29.72 &\textbf{31.78*} \\
 BERT-large-cased & & 17.53 & 24.96 &26.25 &\textbf{27.83*} & & 16.05 & 28.75 &27.88 & \textbf{31.19*} \\
 RoBERTa-base & & 15.05 &28.21 &28.89 &\textbf{30.80*} & &14.17 &33.33 &34.13 &\textbf{34.87*}\\
 RoBERTa-large & & 19.20 & 25.43 &26.76 &\textbf{31.02*} & &17.25 &31.36 &32.39 &\textbf{35.08*} \\ 
\bottomrule
\end{tabular}
\caption{\ac{OOD} fine-tuning results (Micro F1) over six datasets. \(*\) indicates a statistically significant difference (t-test, p\(<\)0.05) when compared to FactMix.}
\label{tab:fine-tining OOD}
\end{table*}

\begin{table}[t]
\centering
\small
\setlength{\tabcolsep}{1mm}
%\resizebox{\linewidth}{!}{
\begin{tabular}{l cccc}
\toprule
  & \multicolumn{4}{c}{\textbf{In-domain Prompt-tuning Results}}        \\ 
\cmidrule{2-5} 
\textbf{Backbone} & \textbf{EntLM} & \textbf{CF} & \textbf{FactMix} & \textbf{PLTR} \\ 
\midrule
% \multirow{4}{*}{\begin{tabular}[c]{@{}c@{}}CoNLL2003 \\ (Dev)\end{tabular}}& BERT-base-cased & 57.98 & 79.78 &83.13 &   \\
% & BERT-large-cased & 69.18 & 83.27 & 85.87 & \\
% & RoBERTa-base &52.44 & 85.81 & 88.51 & \\
% & RoBERTa-large & 68.81 & 88.25 & 89.95 & \\ 
% \hline
BERT-base-cased &54.00 &55.61 & 59.19 &\textbf{63.50*} \\
BERT-large-cased &60.37 &56.49 &60.80 &\textbf{70.46*} \\
\bottomrule
\end{tabular}
%}
\caption{In-domain prompt-tuning results (Micro F1) on CoNLL2003. \(*\) indicates a statistically significant difference (t-test, p\(<\)0.05) when compared to FactMix.}
\label{tab:in-domain prompt-tuning}
\end{table}

\begin{table*}[ht]
\centering
\small
\resizebox{\linewidth}{!}{
\begin{tabular}{l l cccc l cccc}
\toprule
  & & \multicolumn{4}{c}{\textbf{OOD Prompt-tuning Results}} & & \multicolumn{4}{c}{\textbf{OOD Prompt-tuning Results}} \\ 
 \cmidrule{3-6} \cmidrule{8-11} 
 \textbf{Backbone} & \textbf{Dataset} & \textbf{EntLM} & \textbf{CF}     & \textbf{FactMix} & \textbf{PLTR} & \textbf{Dataset} & \textbf{EntLM} & \textbf{CF}     & \textbf{FactMix} & \textbf{PLTR} \\ 
 \midrule
BERT-base-cased & \multirow{2}{*}{TechNews} & 47.16 & 52.36 & 52.44 &\textbf{60.99*} & \multirow{2}{*}{Science} & 15.70 & 18.32 &18.62 &\textbf{20.90*}  \\
 BERT-large-cased & & 52.53 & 48.32 &48.64 &\textbf{61.64*} &  & 15.32 & 15.34 & 16.80 &\textbf{19.77*} \\
 \bottomrule
\end{tabular}}
\caption{\ac{OOD} prompt-tuning results (Micro F1) on TechNews and Science.  \(*\) indicates a statistically significant difference (t-test, p\(<\)0.05) when compared to FactMix.}
\label{tab:prompt-tuning OOD}
\end{table*}

\begin{table*}[ht]
\small
\centering
%\resizebox{\linewidth}{!}{
\begin{tabular}{l ccccc ccccc}
\toprule
 & \multicolumn{5}{c}{\textbf{OOD Fine-tuning Results}} & \multicolumn{5}{c}{\textbf{OOD Prompt-tuning Results}} \\ 
\cmidrule(r){2-6} \cmidrule{7-11} 
\textbf{Dataset} & \textbf{FactMix} & \textbf{NP}  & \textbf{RDW} & \textbf{REW}  & \textbf{PLTR} & \textbf{FactMix} & \textbf{NP}  & \textbf{RDW} & \textbf{REW}  & \textbf{PLTR} \\ 
\midrule
TechNews &65.09 & 66.16 & 66.01 &66.10 &\textbf{67.39} &52.44 &54.01 &55.90 &56.46 & \textbf{60.99}  \\
Science &29.72 &29.84 &30.02 &30.06 & \textbf{31.50} & 18.62 & 18.78 & 18,72 & 19.19 & \textbf{20.90}  \\
\bottomrule
\end{tabular}
%}
\caption{Ablation studies on TechNews and Science.}
\label{tab:ablation}
\end{table*}

To answer RQ1 and RQ2, we assess the performance of \ac{PLTR} on both in-domain and cross-domain few-shot \ac{NER} tasks. 
This evaluation is conducted in two settings: a fine-tuning setting with 100 training instances per type, and a prompt-tuning setting with 5 training instances per type.

\subsection{Results on few-shot fine-tuning (RQ1)}
\label{subs:fine-tuning}
Table~\ref{tab:in-domain fine-tuning} and~\ref{tab:fine-tining OOD} show the in-domain and cross-domain performance in the fine-tuning setting, respectively. 
Based on the results, we have the following observations:
\begin{enumerate*}[label=(\roman*),nosep,leftmargin=*]
\item \ac{PLTR} achieves the highest Micro F1 scores for all datasets and settings, indicating its superior performance.
For instance, when using RoBERTa-large as the backbone, \ac{PLTR} achieves an 88.03\% and 75.14\% F1 score on the CoNLL2003 and TechNews datasets, respectively.
\item \ac{PLTR} significantly outperforms the previous state-of-the-art baselines in both in-domain and cross-domain \ac{NER}. 
For example, \ac{PLTR} exhibits a 1.46\% and 10.64\% improvement over FactMix, on average, on in-domain and cross-domain datasets, respectively.
\item Few-shot cross-domain \ac{NER} is notably more challenging than the in-domain setting, as all methods obtain considerably lower F1 scores.
The performance decay in TechNews is smaller than in other domains, due to its higher overlap with the training set.
\end{enumerate*}
In summary, \ac{PLTR} demonstrates its effectiveness in recognizing named entities from both in-domain and \ac{OOD} examples. 
The use of \acfp{TRF}, along with the incorporation of prompts based on \acp{TRF}, are beneficial for in-domain and cross-domain few-shot \ac{NER}.

\subsection{Results on few-shot prompt-tuning (RQ2)}
\label{subs:prompt-tuning}
To explore the generalizability of \ac{PLTR}, we report in-domain and \ac{OOD} results for the prompt-tuning setting in Table~\ref{tab:in-domain prompt-tuning} and~\ref{tab:prompt-tuning OOD}, respectively. 
We obtain the following insights:
\begin{enumerate*}[label=(\roman*),nosep,leftmargin=*]
\item Due to data sparsity, the overall performance for the prompt-tuning setting is considerably lower than the results of 100-shot fine-tuning.
\item Even with only 5-shot training instances per entity type, \ac{PLTR} achieves the highest performance and outperforms the state-of-the-art baselines by a significant margin, demonstrating the effectiveness and generalizability of \ac{PLTR}.
For example, in the in-domain and cross-domain datasets, \ac{PLTR} achieves an average improvement of 11.58\% and 18.24\% over FactMix, respectively.
\end{enumerate*}
In summary, the \ac{PLTR} framework not only effectively generalizes fine-tuning-based \ac{NER} methods to unseen domains, but also attains the highest F1 scores in the prompt-tuning setting.

\section{Analysis}
\label{sec:analysis}
Now that we have answered our research questions, we take a closer look at \ac{PLTR} to analyze its performance. 
We examine whether the prompts are designed appropriately.
Besides, we study how the number of training samples and selected type-related features influence the performance (Sec.~\ref{sec:features}), how \ac{PLTR} affects representation similarities between the source and target domains, and whether \ac{PLTR} is extensible to different source domains and entity types (Sec.~\ref{sec:source}). 
Furthermore, we provide insights into the possible factors that limit further improvements.

\header{Ablation studies}
To investigate the appropriateness of our prompt design, we conduct ablation studies on few-shot cross-domain \ac{NER} in both fine-tuning and prompt-tuning settings.
The results are presented in Table~\ref{tab:ablation}.
In the ``NP'' variant, prompts are removed during test-time inference.
In this case, the F1 scores across all datasets and settings suffer a significant drop compared to our proposed \ac{PLTR}.
This demonstrates the crucial role of incorporating prompts during both the training and inference processes.
In the ``RDW'' and ``REW'' variants, prompts are constructed using randomly selected words from the source domain and the given example, respectively.
The performance of both the ``RDW'' and ``REW'' model variants consistently falls short of \ac{PLTR}, indicating that \ac{PLTR} effectively identifies important knowledge from the source domain and establishes connections between unseen examples and the knowledge within the source domain.

Additionally, to explore the efficacy of type-related feature selection (refer to Sec.~\ref{subsec:prompt}), we conducted an evaluation of \ac{PLTR} (BERT-base) using various frequency ratios $\rho$ (in Eq.~\ref{eq:MI}).
The results are presented in Table~\ref{tab:frequency ratio}.
As the value of $\rho$ increases, \acp{TRF} extracted using Eq.\ref{eq:MI} become less closely associated with the specified entity type but become more prevalent in other types.
When the value of $\rho$ is raised from 3 to 9, we observed only a slight decrease in the F1 
scores of \ac{PLTR}. 
When the value of $\rho$ is raised to 20, the F1 
score of \ac{PLTR} drops, but still surpasses the state-of-the-art baseline FactMix.
These results indicate that \ac{PLTR} effectively identifies relevant \acp{TRF} for \ac{OOD} examples, considerably mitigating human interventions in the feature extraction process.

\begin{table}[t]
\centering
\small
\resizebox{\linewidth}{!}{
\begin{tabular}{l l ccccc}
\toprule
  & & \multicolumn{5}{c}{\textbf{Frequency ratio $\rho$}}        \\ 
\cmidrule{3-7} 
\textbf{Model} & \textbf{Dataset} & \textbf{3} & \textbf{5} & \textbf{7} & \textbf{9} & \textbf{20}\\ 
\midrule
FactMix &\multirow{2}{*}{AI} &24.67 &24.67 &24.67 &24.67 &24.67\\
PLTR & &\textbf{28.41} &\textbf{26.36} &\textbf{26.61} &\textbf{26.42} &\textbf{25.70} \\
\midrule
FactMix &\multirow{2}{*}{Science} &29.72 &29.72 & 29.72 &29.72 &29.72\\
PLTR & &\textbf{31.78} &\textbf{30.07} &\textbf{30.11} &\textbf{30.58} &\textbf{29.91} \\
\bottomrule
\end{tabular}
}
\caption{Influence of frequency ratio ($\rho$) on AI and Science (BERT-base, fine-tuning).}
\label{tab:frequency ratio}
\end{table}

\begin{figure}
  \centering
  \subfigure[TechNews.]{
    \includegraphics[width=0.48\linewidth]{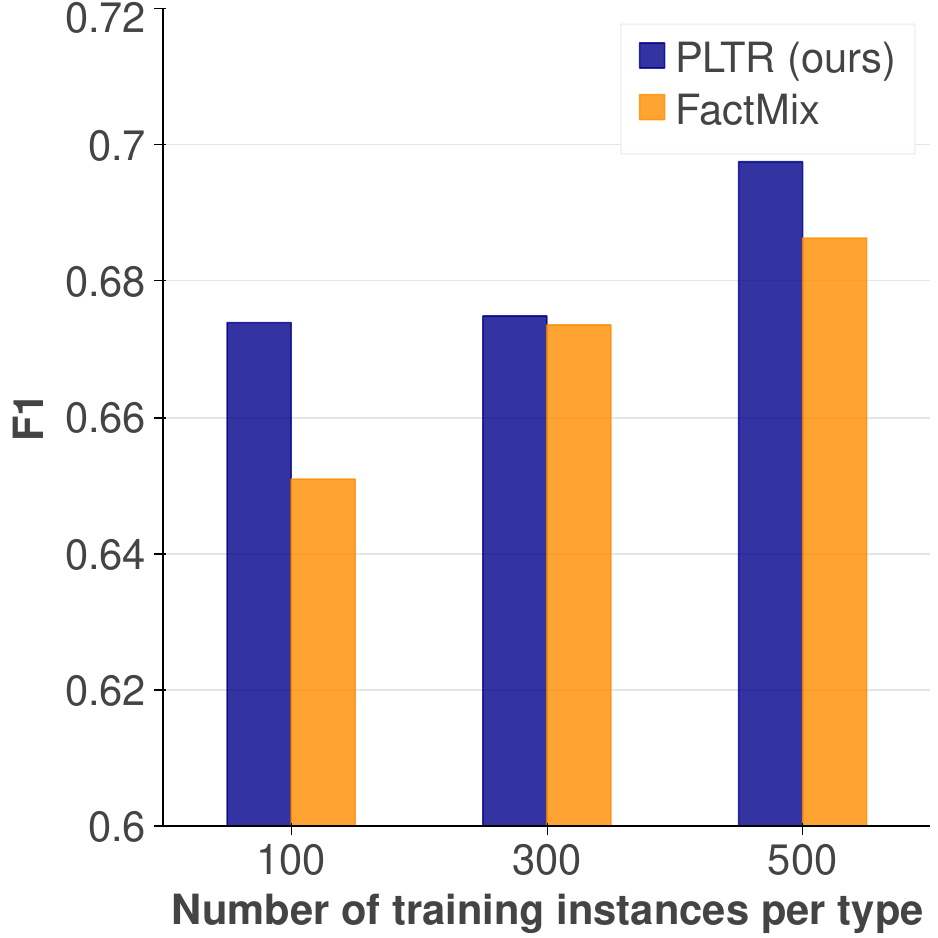}}
   \subfigure[Science.]{
    \includegraphics[width=0.48\linewidth]{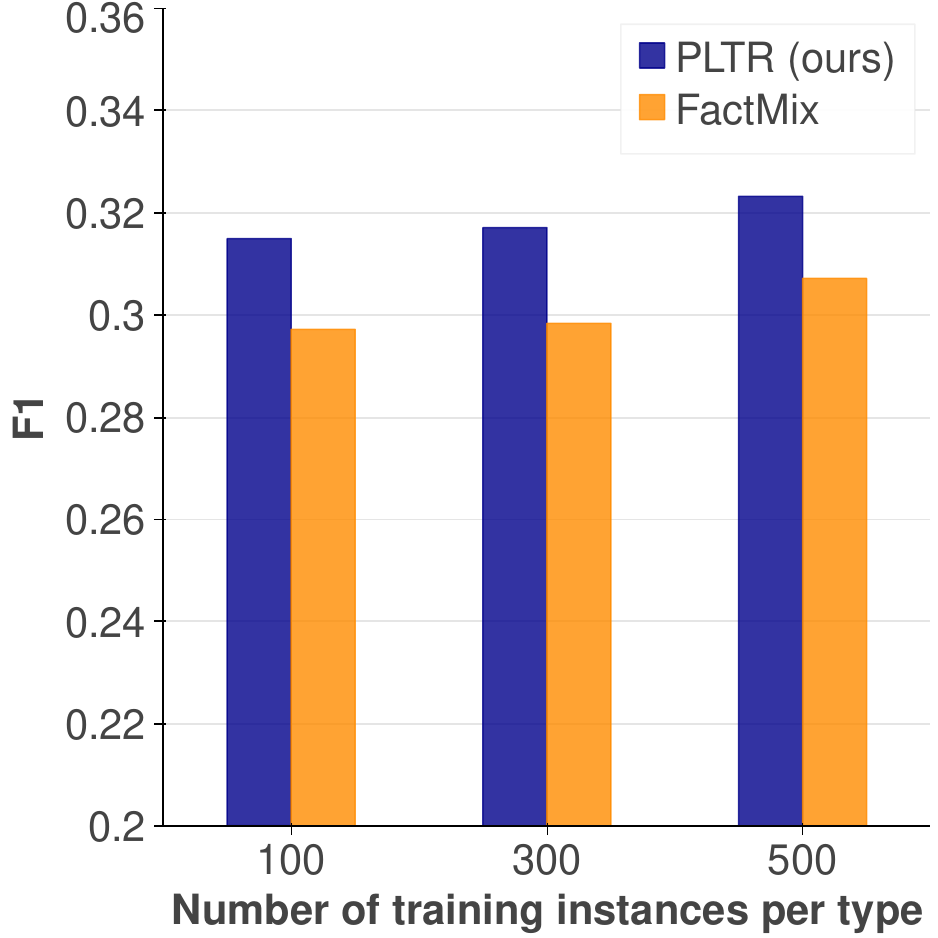}}
    \caption{Influence of training instances on TechNews and Science (BERT-base).}
  \label{fig:training}
\end{figure}

\header{The influence of training samples}
To examine the impact of the number of training samples, we compare the performance of \ac{PLTR} and FactMix on few-shot cross-domain \ac{NER} using 100, 300, and 500 training samples per entity type.
Fig.~\ref{fig:training} displays the results based on the BERT-base-cased model.
\ac{PLTR} exhibits the largest improvements over FactMix when the dataset comprises only 100 training instances per entity type, as opposed to the 300 and 500 training instances scenarios.
Furthermore, \ac{PLTR} consistently outperforms the prior state-of-the-art approach, FactMix, across all experimental settings with varying numbers of training examples, demonstrating its superiority.

\begin{figure}[t]
    \centering
    \includegraphics[width=0.97\linewidth]{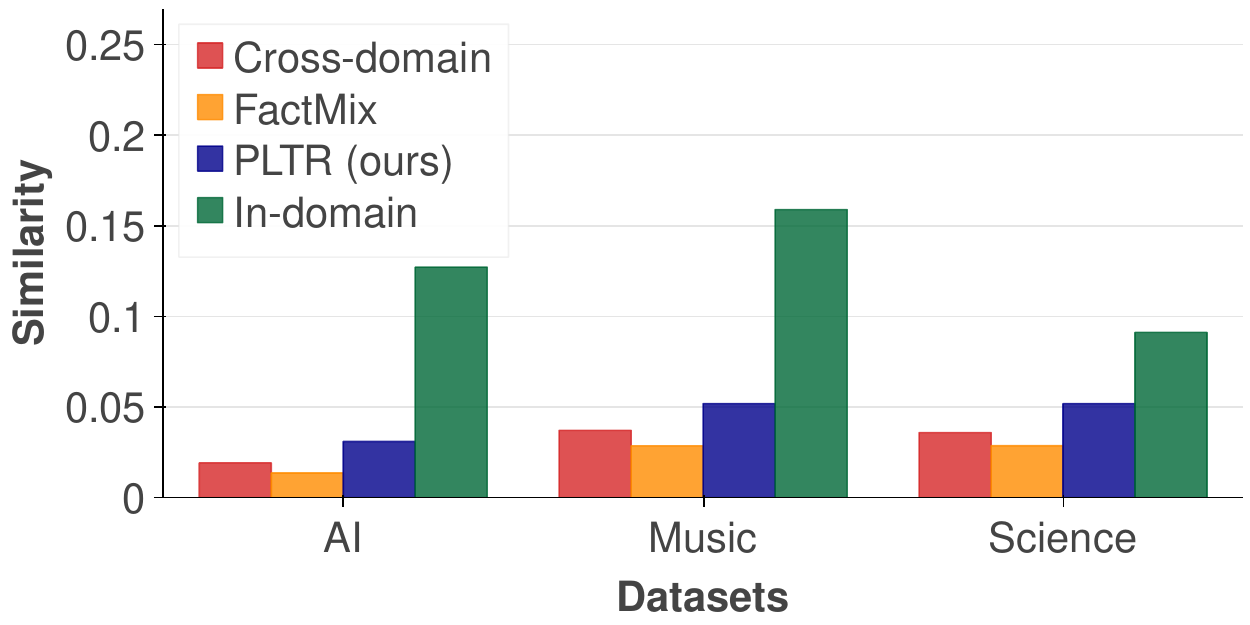}
    \caption{Analysis of sentence similarities on AI, Music, and Science (BERT-base, fine-tuning).}
    \label{fig:similarities}
\end{figure}

\header{Analysis of sentence similarities}
In our analysis of sentence similarities, we investigate the impact of \ac{PLTR} on the representation similarities between the source and target domains.
We compute the average SBERT similarities for sentence representations in \ac{PLTR} (BERT-base) between the source and target domains; the results are presented in Fig.~\ref{fig:similarities}.
With the prompts generated by \ac{PLTR}, the representation similarities between the source and unseen domains noticeably increase.
This is, \ac{PLTR} facilitates a more aligned and connected representation space, mitigating the gap between the source and target domains.

\header{Error analysis}
Although our proposed \ac{PLTR} outperforms state-of-the-art baselines,
we would like to analyze the factors restricting further improvements.
Specifically, we compare the performance of \ac{PLTR} (BERT-base) on sentences of different lengths in the test sets of the CoNLL2003 (In-domain), AI, and Science datasets. The results of the standard fine-tuning setting are provided in Table~\ref{tab:error}.
We observe that the F1 scores of \ac{PLTR} on sentences with more than 35 words (``$> 35$'') are substantially higher than the overall F1 scores.
In contrast, the F1 scores on sentences with 25 to 35 words (``25--35'') or less than 25 words (``$< 25$'') consistently fall below the overall F1 scores.
This suggests that it may be more challenging for \ac{PLTR} to select \acp{TRF} and generate appropriate prompts with less context.

\begin{table}[t]
\centering
\small
\setlength{\tabcolsep}{2mm}
% \resizebox{0\linewidth}{!}{
\begin{tabular}{l cccc}
\toprule
&  \multicolumn{3}{c}{\textbf{Sentence length}} &        \\ 
\cmidrule{2-4} 
\textbf{Dataset} & \textbf{< 25} & \textbf{25--35} & \textbf{> 35} & \textbf{Avg.}\\ 
\midrule
In-domain &80.12 &81.25 & 84.12 & 82.05 \\
AI &25.86 &24.71 &29.65 & 28.41\\
Science &23.86 &29.62 & 32.87 & 31.78 \\
\bottomrule
\end{tabular}
\caption{Error analysis on sentence lengths in test sets (BERT-base, fine-tuning).}
\label{tab:error}
\end{table}

\section{Conclusions}
In this paper, we establish a new state-of-the-art framework, \ac{PLTR}, for few-shot cross-domain \ac{NER}.
To capture useful knowledge from the source domain, \ac{PLTR} employs mutual information criteria to extract type-related features.
\ac{PLTR} automatically selects pertinent features and generates a unique prompt for each unseen example, bridging the gap between domains.  
Experimental results show that \ac{PLTR} not only effectively generalizes standard fine-tuning methods to unseen domains, but also demonstrates promising performance when incorporated with prompt-tuning-based approaches.
Additionally, \ac{PLTR} substantially narrows the disparity between in-domain examples and \ac{OOD} instances, enhancing the similarities of their sentence representations. 

\clearpage

\section*{Limitations}
While \ac{PLTR} achieves a new state-of-the-art performance, it has several limitations. First, the number of type-related features for prompt construction needs to be manually preset. Second, \ac{PLTR} relies on identifying \acp{TRF}, which are tokens strongly associated with entity types. Extracting and incorporating more complex features, such as phrases, represents a promising direction for future research. In the future, we also plan to incorporate \ac{PLTR} with different kinds of pre-trained language models, such as autoregressive language models.

\section*{Ethics statement}
The paper presents a prompt-based method for recognizing named entities in unseen domains with limited labeled in-domain examples. 
However, the constructed prompts and model-predicted results still have a considerable amount of misinformation.
Besides, the reliance on black-box pre-trained language models raises concerns. 
Hence, caution and further research are required prior to deploying this method in real-world applications.

\section*{Acknowledgement}
This work was supported by the National Key R\&D Program of China (2020YFB1406704, 2022YFC3303004), the Natural Science Foundation of China (62272274, 61972234, 62072279, 62102234, 62202271), the Natural Science Foundation of Shandong Province (ZR2021QF129, ZR2022QF004), the Key Scientific and Technological Innovation Program of Shandong Province (2019JZZY010129), the Fundamental Research Funds of Shandong University, the China Scholarship Council under grant nr. 202206220085,
the Hybrid Intelligence Center, a 10-year program funded by the Dutch Ministry of Education, Culture and Science through the Netherlands Organization for Scientific Research, \url{https://hybrid-intelligence-centre.nl}, and project LESSEN with project number NWA.1389.20.183 of the research program NWA ORC 2020/21, which is (partly) financed by the Dutch Research Council (NWO).

\bibliography{custom}

\begin{thebibliography}{47}
\expandafter\ifx\csname natexlab\endcsname\relax\def\natexlab#1{#1}\fi

\bibitem[{Baxter(2000)}]{baxter2000model}
Jonathan Baxter. 2000.
\newblock A model of inductive bias learning.
\newblock \emph{J. Artif. Intell. Res.}, 12:149--198.

\bibitem[{Ben{-}David et~al.(2022)Ben{-}David, Oved, and Reichart}]{DBLP:journals/tacl/Ben-DavidOR22}
Eyal Ben{-}David, Nadav Oved, and Roi Reichart. 2022.
\newblock {PADA:} example-based prompt learning for on-the-fly adaptation to unseen domains.
\newblock \emph{Trans. Assoc. Comput. Linguistics}, 10:414--433.

\bibitem[{Brown et~al.(2020)Brown, Mann, Ryder, Subbiah, Kaplan, Dhariwal, Neelakantan, Shyam, Sastry, Askell, Agarwal, Herbert{-}Voss, Krueger, Henighan, Child, Ramesh, Ziegler, Wu, Winter, Hesse, Chen, Sigler, Litwin, Gray, Chess, Clark, Berner, McCandlish, Radford, Sutskever, and Amodei}]{DBLP:conf/nips/BrownMRSKDNSSAA20}
Tom~B. Brown, Benjamin Mann, Nick Ryder, Melanie Subbiah, Jared Kaplan, Prafulla Dhariwal, Arvind Neelakantan, Pranav Shyam, Girish Sastry, Amanda Askell, Sandhini Agarwal, Ariel Herbert{-}Voss, Gretchen Krueger, Tom Henighan, Rewon Child, Aditya Ramesh, Daniel~M. Ziegler, Jeffrey Wu, Clemens Winter, Christopher Hesse, Mark Chen, Eric Sigler, Mateusz Litwin, Scott Gray, Benjamin Chess, Jack Clark, Christopher Berner, Sam McCandlish, Alec Radford, Ilya Sutskever, and Dario Amodei. 2020.
\newblock Language models are few-shot learners.
\newblock In \emph{NeurIPS}.

\bibitem[{Chen et~al.(2022{\natexlab{a}})Chen, Liu, Lin, Han, and Sun}]{chen2022few}
Jiawei Chen, Qing Liu, Hongyu Lin, Xianpei Han, and Le~Sun. 2022{\natexlab{a}}.
\newblock Few-shot named entity recognition with self-describing networks.
\newblock In \emph{ACL}, pages 5711--5722.

\bibitem[{Chen et~al.(2021)Chen, Aguilar, Neves, and Solorio}]{chen2021data}
Shuguang Chen, Gustavo Aguilar, Leonardo Neves, and Thamar Solorio. 2021.
\newblock Data augmentation for cross-domain named entity recognition.
\newblock In \emph{EMNLP}, pages 5346--5356.

\bibitem[{Chen et~al.(2022{\natexlab{b}})Chen, Zheng, and Yang}]{DBLP:journals/corr/abs-2211-04337}
Yanru Chen, Yanan Zheng, and Zhilin Yang. 2022{\natexlab{b}}.
\newblock Prompt-based metric learning for few-shot {NER}.
\newblock \emph{CoRR}, abs/2211.04337.

\bibitem[{Cui et~al.(2021)Cui, Wu, Liu, Yang, and Zhang}]{cui-etal-2021-template}
Leyang Cui, Yu~Wu, Jian Liu, Sen Yang, and Yue Zhang. 2021.
\newblock Template-based named entity recognition using {BART}.
\newblock In \emph{ACL-IJCNLP}, pages 1835--1845.

\bibitem[{Cui and Zhang(2019)}]{lan}
Leyang Cui and Yue Zhang. 2019.
\newblock Hierarchically-refined label attention network for sequence labeling.
\newblock In \emph{EMNLP-IJCNLP}, pages 4115--4128.

\bibitem[{Das et~al.(2022)Das, Katiyar, Passonneau, and Zhang}]{das2021container}
Sarkar Snigdha~Sarathi Das, Arzoo Katiyar, Rebecca~J. Passonneau, and Rui Zhang. 2022.
\newblock Container: Few-shot named entity recognition via contrastive learning.
\newblock In \emph{ACL}, pages 6338--6353.

\bibitem[{Devlin et~al.(2019)Devlin, Chang, Lee, and Toutanova}]{devlin2018bert}
Jacob Devlin, Ming{-}Wei Chang, Kenton Lee, and Kristina Toutanova. 2019.
\newblock {BERT:} pre-training of deep bidirectional transformers for language understanding.
\newblock In \emph{NAACL-HLT}, pages 4171--4186.

\bibitem[{Dong et~al.(2023)Dong, Wang, Zhao, Zhao, Guo, Fu, Hui, Zeng, He, Li, Wang, Cui, and Xu}]{DBLP:journals/corr/abs-2308-14533}
Guanting Dong, Zechen Wang, Jinxu Zhao, Gang Zhao, Daichi Guo, Dayuan Fu, Tingfeng Hui, Chen Zeng, Keqing He, Xuefeng Li, Liwen Wang, Xinyue Cui, and Weiran Xu. 2023.
\newblock A multi-task semantic decomposition framework with task-specific pre-training for few-shot {NER}.
\newblock \emph{CoRR}, abs/2308.14533.

\bibitem[{Fang et~al.(2023)Fang, Wang, Meng, Xie, Huang, and Jiang}]{DBLP:conf/acl/FangWMXHJ23}
Jinyuan Fang, Xiaobin Wang, Zaiqiao Meng, Pengjun Xie, Fei Huang, and Yong Jiang. 2023.
\newblock {MANNER:} {A} variational memory-augmented model for cross domain few-shot named entity recognition.
\newblock In \emph{ACL}, pages 4261--4276.

\bibitem[{Gutierrez et~al.(2022)Gutierrez, McNeal, Washington, Chen, Li, Sun, and Su}]{DBLP:conf/emnlp/GutierrezMWCLS022}
Bernal~Jimenez Gutierrez, Nikolas McNeal, Clayton Washington, You Chen, Lang Li, Huan Sun, and Yu~Su. 2022.
\newblock Thinking about {GPT-3} in-context learning for biomedical {IE}? think again.
\newblock In \emph{EMNLP}, pages 4497--4512.

\bibitem[{Han et~al.(2023)Han, Peng, Yang, Wang, Liu, and Wan}]{DBLP:journals/corr/abs-2305-14450}
Ridong Han, Tao Peng, Chaohao Yang, Benyou Wang, Lu~Liu, and Xiang Wan. 2023.
\newblock Is information extraction solved by {ChatGPT}? {An} analysis of performance, evaluation criteria, robustness and errors.
\newblock \emph{CoRR}, abs/2305.14450.

\bibitem[{Hu et~al.(2023)Hu, Guo, Liu, Li, Chen, Wan, and Chang}]{DBLP:conf/aaai/HuGLLCWC23}
Jinpeng Hu, Dandan Guo, Yang Liu, Zhuo Li, Zhihong Chen, Xiang Wan, and Tsung{-}Hui Chang. 2023.
\newblock A simple yet effective subsequence-enhanced approach for cross-domain {NER}.
\newblock In \emph{AAAI}, pages 12890--12898.

\bibitem[{Huang et~al.(2021)Huang, Li, Subudhi, Jose, Balakrishnan, Chen, Peng, Gao, and Han}]{huang-etal-2021-shot}
Jiaxin Huang, Chunyuan Li, Krishan Subudhi, Damien Jose, Shobana Balakrishnan, Weizhu Chen, Baolin Peng, Jianfeng Gao, and Jiawei Han. 2021.
\newblock Few-shot named entity recognition: An empirical baseline study.
\newblock In \emph{EMNLP}, pages 10408--10423.

\bibitem[{Jia et~al.(2019)Jia, Liang, and Zhang}]{jia2019cross}
Chen Jia, Xiaobo Liang, and Yue Zhang. 2019.
\newblock Cross-domain {NER} using cross-domain language modeling.
\newblock In \emph{ACL}, pages 2464--2474.

\bibitem[{Jia and Zhang(2020)}]{jia2020multi}
Chen Jia and Yue Zhang. 2020.
\newblock Multi-cell compositional {LSTM} for {NER} domain adaptation.
\newblock In \emph{ACL}, pages 5906--5917.

\bibitem[{Lee et~al.(2022)Lee, Kadakia, Tan, Agarwal, Feng, Shibuya, Mitani, Sekiya, Pujara, and Ren}]{DBLP:conf/acl/LeeKTAF0MSPR22}
Dong{-}Ho Lee, Akshen Kadakia, Kangmin Tan, Mahak Agarwal, Xinyu Feng, Takashi Shibuya, Ryosuke Mitani, Toshiyuki Sekiya, Jay Pujara, and Xiang Ren. 2022.
\newblock Good examples make a faster learner: Simple demonstration-based learning for low-resource {NER}.
\newblock In \emph{ACL}, pages 2687--2700.

\bibitem[{Lee et~al.(2018)Lee, Dernoncourt, and Szolovits}]{lee-etal-2018-transfer}
Ji~Young Lee, Franck Dernoncourt, and Peter Szolovits. 2018.
\newblock Transfer learning for named-entity recognition with neural networks.
\newblock In \emph{LREC}.

\bibitem[{Li et~al.(2022)Li, Hu, and Chen}]{DBLP:conf/coling/LiHC22}
Dongfang Li, Baotian Hu, and Qingcai Chen. 2022.
\newblock Prompt-based text entailment for low-resource named entity recognition.
\newblock In \emph{COLING}, pages 1896--1903.

\bibitem[{Liu et~al.(2019)Liu, Ott, Goyal, Du, Joshi, Chen, Levy, Lewis, Zettlemoyer, and Stoyanov}]{liu2019roberta}
Yinhan Liu, Myle Ott, Naman Goyal, Jingfei Du, Mandar Joshi, Danqi Chen, Omer Levy, Mike Lewis, Luke Zettlemoyer, and Veselin Stoyanov. 2019.
\newblock Roberta: {A} robustly optimized {BERT} pretraining approach.
\newblock \emph{CoRR}, abs/1907.11692.

\bibitem[{Liu et~al.(2021)Liu, Xu, Yu, Dai, Ji, Cahyawijaya, Madotto, and Fung}]{liu2021crossner}
Zihan Liu, Yan Xu, Tiezheng Yu, Wenliang Dai, Ziwei Ji, Samuel Cahyawijaya, Andrea Madotto, and Pascale Fung. 2021.
\newblock {CrossNER}: Evaluating cross-domain named entity recognition.
\newblock In \emph{AAAI}, volume~35, pages 13452--13460.

\bibitem[{Loshchilov and Hutter(2019)}]{DBLP:conf/iclr/LoshchilovH19}
Ilya Loshchilov and Frank Hutter. 2019.
\newblock Decoupled weight decay regularization.
\newblock In \emph{ICLR}.

\bibitem[{Ma et~al.(2023)Ma, Lin, Chen, Zhou, Wang, Gui, Zhang, Gao, and Chen}]{DBLP:conf/acl/MaLCZWGZGC23}
Ruotian Ma, Zhang Lin, Xuanting Chen, Xin Zhou, Junzhe Wang, Tao Gui, Qi~Zhang, Xiang Gao, and Yun~Wen Chen. 2023.
\newblock Coarse-to-fine few-shot learning for named entity recognition.
\newblock In \emph{ACL}, pages 4115--4129.

\bibitem[{Ma et~al.(2022)Ma, Zhou, Gui, Tan, Li, Zhang, and Huang}]{ma2021template}
Ruotian Ma, Xin Zhou, Tao Gui, Yiding Tan, Linyang Li, Qi~Zhang, and Xuanjing Huang. 2022.
\newblock Template-free prompt tuning for few-shot {NER}.
\newblock In \emph{NAACL}, pages 5721--5732.

\bibitem[{Ma and Hovy(2016)}]{lstm-cnn-crf}
Xuezhe Ma and Eduard Hovy. 2016.
\newblock End-to-end sequence labeling via bi-directional {LSTM}-{CNN}s-{CRF}.
\newblock In \emph{ACL}, pages 1064--1074.

\bibitem[{Nadeau and Sekine(2007)}]{nadeau2007survey}
David Nadeau and Satoshi Sekine. 2007.
\newblock A survey of named entity recognition and classification.
\newblock \emph{Lingvisticae Investigationes}, 30(1):3--26.

\bibitem[{OpenAI(2023)}]{DBLP:journals/corr/abs-2303-08774}
OpenAI. 2023.
\newblock {GPT-4} technical report.
\newblock \emph{CoRR}, abs/2303.08774.

\bibitem[{Pan et~al.(2013)Pan, Toh, and Su}]{pan2013transfer}
Sinno~Jialin Pan, Zhiqiang Toh, and Jian Su. 2013.
\newblock Transfer joint embedding for cross-domain named entity recognition.
\newblock \emph{{ACM} Trans. Inf. Syst.}, 31(2):1--27.

\bibitem[{Reimers and Gurevych(2019)}]{DBLP:conf/emnlp/ReimersG19}
Nils Reimers and Iryna Gurevych. 2019.
\newblock Sentence-{BERT}: Sentence embeddings using siamese {BERT}-networks.
\newblock In \emph{EMNLP-IJCNLP}, pages 3980--3990.

\bibitem[{Sang and Meulder(2003)}]{sang2003introduction}
Erik F. Tjong~Kim Sang and Fien~De Meulder. 2003.
\newblock Introduction to the {CoNLL-2003} shared task: Language-independent named entity recognition.
\newblock In \emph{CoNLL}, pages 142--147.

\bibitem[{Shen et~al.(2021)Shen, Liu, He, Zhang, Xu, Yu, and Cui}]{Shen2021TowardsOG}
Zheyan Shen, Jiashuo Liu, Yue He, Xingxuan Zhang, Renzhe Xu, Han Yu, and Peng Cui. 2021.
\newblock Towards out-of-distribution generalization: {A} survey.
\newblock \emph{CoRR}, abs/2108.13624.

\bibitem[{Sun et~al.(2023)Sun, Dong, Li, Wan, Wang, Zhang, Li, Cheng, Lyu, Wu, and Wang}]{DBLP:journals/corr/abs-2306-09719}
Xiaofei Sun, Linfeng Dong, Xiaoya Li, Zhen Wan, Shuhe Wang, Tianwei Zhang, Jiwei Li, Fei Cheng, Lingjuan Lyu, Fei Wu, and Guoyin Wang. 2023.
\newblock Pushing the limits of {ChatGPT} on {NLP} tasks.
\newblock \emph{CoRR}, abs/2306.09719.

\bibitem[{Wang et~al.(2022)Wang, Chen, Fan, Sun, Tao, Hou, Wang, Yang, Zhou, Guo, Qi, Wu, Li, Nakamura, Ye, Savvides, Raj, Shinozaki, Schiele, Wang, Xie, and Zhang}]{wangusb}
Yidong Wang, Hao Chen, Yue Fan, Wang Sun, Ran Tao, Wenxin Hou, Renjie Wang, Linyi Yang, Zhi Zhou, Lan{-}Zhe Guo, Heli Qi, Zhen Wu, Yu{-}Feng Li, Satoshi Nakamura, Wei Ye, Marios Savvides, Bhiksha Raj, Takahiro Shinozaki, Bernt Schiele, Jindong Wang, Xing Xie, and Yue Zhang. 2022.
\newblock {USB:} {A} unified semi-supervised learning benchmark for classification.
\newblock In \emph{NeurIPS}.

\bibitem[{Wang et~al.(2023)Wang, Chen, Heng, Hou, Fan, Wu, Wang, Savvides, Shinozaki, Raj, Schiele, and Xie}]{wang2022freematch}
Yidong Wang, Hao Chen, Qiang Heng, Wenxin Hou, Yue Fan, Zhen Wu, Jindong Wang, Marios Savvides, Takahiro Shinozaki, Bhiksha Raj, Bernt Schiele, and Xing Xie. 2023.
\newblock Freematch: Self-adaptive thresholding for semi-supervised learning.
\newblock In \emph{ICLR}.

\bibitem[{Wang et~al.(2021)Wang, Song, Ren, Ren, Chen, Liu, Li, and de~Rijke}]{DBLP:conf/sigir/WangSRRCLLR21}
Zihan Wang, Hongye Song, Zhaochun Ren, Pengjie Ren, Zhumin Chen, Xiaozhong Liu, Hongsong Li, and Maarten de~Rijke. 2021.
\newblock Cross-domain contract element extraction with a bi-directional feedback clause-element relation network.
\newblock In \emph{SIGIR}, pages 1003--1012.

\bibitem[{Weischedel et~al.(2013)Weischedel, Palmer, Marcus, Hovy, Pradhan, Ramshaw, Xue, Taylor, Kaufman, Franchini, El-Bachouti, Belvin, and Houston}]{weischedel2013ontonotes}
Ralph Weischedel, Martha Palmer, Mitchell Marcus, Eduard Hovy, Sameer Pradhan, Lance Ramshaw, Nianwen Xue, Ann Taylor, Jeff Kaufman, Michelle Franchini, Mohammed El-Bachouti, Robert Belvin, and Ann Houston. 2013.
\newblock {OntoNotes} release 5.0.
\newblock Linguistic Data Consortium, Philadelphia, PA.
\newblock LDC2013T19.

\bibitem[{Wiseman and Stratos(2019)}]{label-agnostic}
Sam Wiseman and Karl Stratos. 2019.
\newblock Label-agnostic sequence labeling by copying nearest neighbors.
\newblock In \emph{ACL}, pages 5363--5369.

\bibitem[{Yamada et~al.(2020)Yamada, Asai, Shindo, Takeda, and Matsumoto}]{luke}
Ikuya Yamada, Akari Asai, Hiroyuki Shindo, Hideaki Takeda, and Yuji Matsumoto. 2020.
\newblock {LUKE}: Deep contextualized entity representations with entity-aware self-attention.
\newblock In \emph{EMNLP}, pages 6442--6454.

\bibitem[{Yang et~al.(2022)Yang, Yuan, Cui, Gao, and Zhang}]{DBLP:conf/coling/YangYCGZ22}
Linyi Yang, Lifan Yuan, Leyang Cui, Wenyang Gao, and Yue Zhang. 2022.
\newblock Factmix: Using a few labeled in-domain examples to generalize to cross-domain named entity recognition.
\newblock In \emph{COLING}, pages 5360--5371.

\bibitem[{Yang and Katiyar(2020)}]{nearest-neighbor-crf}
Yi~Yang and Arzoo Katiyar. 2020.
\newblock Simple and effective few-shot named entity recognition with structured nearest neighbor learning.
\newblock In \emph{EMNLP}, pages 6365--6375.

\bibitem[{Yang et~al.(2017)Yang, Salakhutdinov, and Cohen}]{jointNER}
Zhilin Yang, Ruslan Salakhutdinov, and William~W. Cohen. 2017.
\newblock Transfer learning for sequence tagging with hierarchical recurrent networks.
\newblock In \emph{ICLR}.

\bibitem[{Zeng et~al.(2020)Zeng, Li, Zhai, and Zhang}]{zeng-etal-2020-counterfactual}
Xiangji Zeng, Yunliang Li, Yuchen Zhai, and Yin Zhang. 2020.
\newblock Counterfactual generator: A weakly-supervised method for named entity recognition.
\newblock In \emph{EMNLP}, pages 7270--7280.

\bibitem[{Zhang et~al.(2022)Zhang, Yu, Wang, Liu, Su, and Xu}]{DBLP:conf/sigir/00010WLSX22}
Xinghua Zhang, Bowen Yu, Yubin Wang, Tingwen Liu, Taoyu Su, and Hongbo Xu. 2022.
\newblock Exploring modular task decomposition in cross-domain named entity recognition.
\newblock In \emph{SIGIR}, pages 301--311.

\bibitem[{Zhang et~al.(2020)Zhang, Yu, Cui, Wu, Wen, and Wang}]{zhang2020every}
Yufeng Zhang, Xueli Yu, Zeyu Cui, Shu Wu, Zhongzhen Wen, and Liang Wang. 2020.
\newblock Every document owns its structure: Inductive text classification via graph neural networks.
\newblock In \emph{ACL}, pages 334--339.

\bibitem[{Zheng et~al.(2022)Zheng, Chen, and Ma}]{DBLP:conf/acl/ZhengCM22}
Junhao Zheng, Haibin Chen, and Qianli Ma. 2022.
\newblock Cross-domain named entity recognition via graph matching.
\newblock In \emph{ACL}, pages 2670--2680.

\end{thebibliography}
\bibliographystyle{acl_natbib}

\appendix

\section{Appendix}

\begin{table*}[ht]
\small
\centering
\resizebox{\linewidth}{!}{
\begin{tabular}{l l l ccccc ccccc}
\toprule
& & & \multicolumn{5}{c}{\textbf{Source: CoNLL2003}} & \multicolumn{5}{c}{\textbf{Source: OntoNotes}} \\ 
\cmidrule(r){4-8} \cmidrule{9-13} 
\textbf{Setting}& \textbf{Model} & \textbf{Dataset} & \textbf{PER} & \textbf{LOC}  & \textbf{ORG} & \textbf{MISC}  & \textbf{Avg.} & \textbf{PER} & \textbf{LOC}  & \textbf{ORG} & \textbf{EVENT}  & \textbf{Avg.} \\ 
\midrule
\multirow{4}{*}{\begin{tabular}[l]{@{}l@{}}OOD\\ Fine-tuning \\ Results\\\end{tabular}}& FactMix &\multirow{2}{*}{TechNews} & 85.65 & 59.45 & 59.31 & 24.66 & 65.20 & 56.96 & 16.06 & 41.84 & -- & 44.11  \\
& PLTR & & \textbf{86.00} & \textbf{71.34} & \textbf{59.93} & \textbf{26.25} & \textbf{67.39}  & \textbf{86.14} & \textbf{18.33} & \textbf{57.71} & -- & \textbf{65.31}  \\
\cmidrule{2-13} 
& FactMix&\multirow{2}{*}{Science} & 35.43 & 31.28 & 24.46 & 23.58 & 29.72 & 13.60 & 15.84 & 22.66 & 3.77 & 17.15  \\
& PLTR & & \textbf{36.27} & \textbf{39.38} & \textbf{36.57} & \textbf{29.99} & \textbf{31.78} & \textbf{38.51} & \textbf{16.79} & \textbf{23.20} & \textbf{7.61} & \textbf{27.46}   \\
\midrule
\multirow{4}{*}{\begin{tabular}[l]{@{}l@{}}OOD\\ Prompt-tuning \\ Results\\\end{tabular}}& FactMix &\multirow{2}{*}{TechNews} & 82.88 & 55.05 & 39.82 & 16.12 & 52.44 & 77.66 & 16.20 & 38.92 & -- & 53.19    \\
& PLTR & & \textbf{87.91}& \textbf{56.73} & \textbf{42.69} & \textbf{29.21} & \textbf{60.99}& \textbf{79.27} & \textbf{20.26} & \textbf{42.67} & -- & \textbf{54.91}   \\
\cmidrule{2-13}
& FactMix&\multirow{2}{*}{Science} & 29.87 & 23.17 & 8.47 & 14.30 & 18.62 & 33.14 & 3.96 & 7.54 & 2.90 & 19.83 \\
& PLTR& & \textbf{38.51} & \textbf{23.19} & \textbf{10.46} & \textbf{14.33} & \textbf{20.90}& \textbf{34.98} & \textbf{11.83} & \textbf{17.27} & \textbf{6.67} & \textbf{22.32}   \\
\bottomrule
\end{tabular}
}
\caption{Influence of source domains (BERT-base). In TechNews, there are no annotations for "EVENT" entities.}
\label{tab:source}
\end{table*}

\subsection{Training algorithm of \ac{PLTR}}
\label{subsec:training}
Algorithm~\ref{alg:training} gives the detailed training algorithm of \ac{PLTR}. To start, we establish a basic model $\mathcal{M}_{b}$ based on Pre-trained Language Models (PLM) and initialize its parameters $\mathbf{\Theta}$ (lines 1-2).
To capture knowledge from the source domain, \ac{PLTR} identifies type-related features using mutual information criteria (line 3).
Next, given an input sentence $\mathbf{x}\in\mathcal{D}_\mathit{in}$, \ac{PLTR} automatically selects relevant \acp{TRF} $\mathcal{R}'(\mathbf{x})\subseteq\mathcal{R}$ by formulating the selection process as a cloze-style task for $\mathcal{M}_{b}$ (line 7).
Furthermore, to incorporate entity type information into prompts, \ac{PLTR} constructs a unique prompt $f'(\mathbf{x})$ for each input $\mathbf{x}$, and these prompts are then fed into $\mathcal{M}_{b}$ for entity recognition (lines 8-9).
Finally, we iteratively refine the parameters $\mathbf{\Theta}$ by jointly optimizing two loss functions: the \ac{NER} task loss function $\mathcal{L}'_\mathit{NER}$ and the \ac{TRF} selection loss function $\mathcal{L}_\mathit{gen}$ (line 11).  
Note that, during inference, \ac{PLTR} generates a unique prompt for each sentence within the unseen target domain using extracted \acp{TRF} $\mathcal{R}$. In this way, knowledge from the source domain is explicitly integrated into both the training and inference phases. 
\begin{algorithm}[t]
    \caption{Training Algorithm for \ac{PLTR}.}
    \label{alg:training}      
    \begin{algorithmic}[1] 
    \Require The source dataset $\mathcal{D}_{in}$; the basic model $\mathcal{M}_{b}$ with parameters $\mathbf{\Theta}$; the frequency ratio $\rho$; the number of selected type-related features $K$; the loss weight $\alpha$; the number of epochs $\mathit{epoch}$.
    \Ensure The extracted type-related features $\mathcal{R}$ and the trained basic model $\mathcal{M}'_{b}$;
    \State Establish the basic model $\mathcal{M}_{b}$;
    \State Initialize model parameters $\mathbf{\Theta}$;
    \State Extract type-related features $\mathcal{R}$ for all entity types from the source dataset $\mathcal{D}_\mathit{in}$ (Eq.~\ref{eq:MI});
    \While{$i \leq \mathit{epoch}$ }
    \For{Sample a batch $\mathcal{X} \subseteq \mathcal{D}_\mathit{in}$}
    \For{all sentences $\mathbf{x} \in \mathcal{X}$}
    \State Select relevant \acp{TRF} $\mathcal{R}'(\mathbf{x})$ for in-
    \State put $\mathbf{x}$ (Eq.~\ref{eq:prompt1} and~\ref{eq:select});
    \State Transform $\mathbf{x}$ into the prompt tem-
    \State plate $f'(\mathbf{x})$ (Eq.~\ref{eq:prompt2});
    \State Input $f'(\mathbf{x})$ into $\mathcal{M}_{b}$ for prediction;
    \EndFor
    \State Update $\mathbf{\Theta}$ by optimizing $\mathcal{L}$ (Eq.~\ref{eq:overall});
    \EndFor
    \EndWhile
    \end{algorithmic}
\end{algorithm}

\subsection{Influence of the number of selected type-related features}
\label{sec:features}
\begin{figure}[t]
    \centering
    \includegraphics[width=\linewidth]{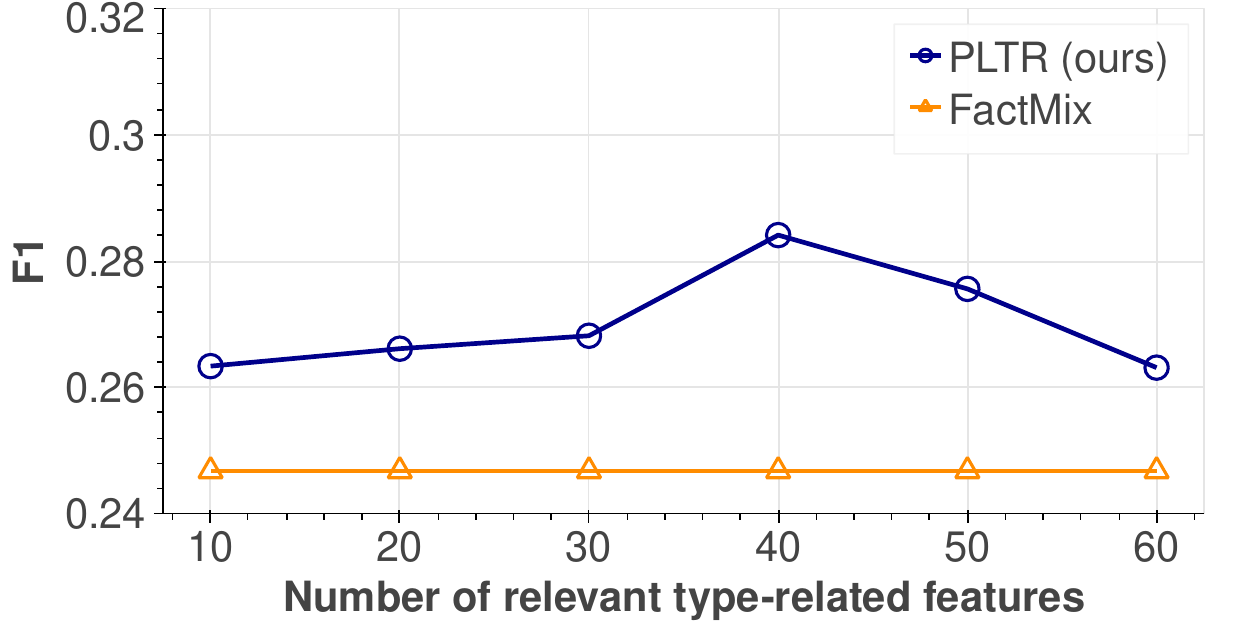}
    \caption{Influence of the number of selected relevant type-related features ($K$) on AI (BERT-base, fine-tuning).}
    \label{fig:TRF}
\end{figure}

We evaluate \ac{PLTR} based on BERT-base in fine-tuning setting, with the number of selected relevant type-related features $K$ varying from 10 to 60.
The results are shown in Fig.~\ref{fig:TRF}.
Our observations indicate that as the number of type-related features increases, the performance (F1 score) of \ac{PLTR} initially improves because the model incorporated with more features is able to encode more useful knowledge from the source domain.
But notice that the performance drops when the number of type-related features is too large.
In our experiments, we set the number of type-related features to 40 on all datasets.

\subsection{Influence of source domains}
\label{sec:source}
We explore the performance of our proposed \ac{PLTR} when trained on data from different source domains, i.e., CoNLL2003 and OntoNotes.
Results in both the fine-tuning and prompt-tuning settings are shown in Table~\ref{tab:source}. 
Our observations indicate that our proposed \ac{PLTR} consistently outperforms FactMix when trained on different domains.
For instance, \ac{PLTR} achieves an average improvement of 5.42\% and 4.22\% over FactMix for "LOC" entities when using CoNLL2003 and OntoNotes as source datasets, respectively. 
This highlights \ac{PLTR}'s capacity to extend to various source domains and entity types.

\end{document}